
\def\lastrev { To appear in {\it Nature} }

\magnification=1100

\hsize = 6.5truein  
\vsize = 9.5truein

\baselineskip 14pt  

\ifnum\mag=1100
\font\bigb=cmbx10 scaled 1308  
\font\vbig=cmbx10 scaled 1571  
\else
\font\bigb=cmbx10 scaled\magstep1
\font\vbig=cmbx10 scaled\magstep2
\fi

\def\etal{\it et al. \rm}

\def\ten#1{\times 10^{#1}}
\def\msun { \rm {M_\odot}}
\def\angs {{ \rm \AA }}
%
%
\def\spose#1{\hbox to 0pt{#1\hss}}
\def\simlt{\mathrel{\spose{\lower 3pt\hbox{$\mathchar"218$}}
     \raise 2.0pt\hbox{$\mathchar"13C$}}}
\def\simgt{\mathrel{\spose{\lower 3pt\hbox{$\mathchar"218$}}
     \raise 2.0pt\hbox{$\mathchar"13E$}}}

\def\ref{\par\hangindent=1cm\hangafter=1\noindent}
\def\fig#1 {\parindent 0pt\hangindent=8em ${}$ \hbox to 8em {#1\hfil}$\!$}

\raggedbottom

\def\sect#1{ \par\noindent \line{} \vskip 0.1in plus 0.4in \goodbreak
      \line{ {\bigb #1 \hfil}} \nobreak\medskip}

\centerline {\vbig Possible gravitational microlensing of a star }
\vskip 10pt
\centerline{\vbig in the Large Magellanic Cloud}
\vskip 0.3truein

{\bf
\centerline {C.  Alcock$^{\ast,\dagger}$, C.W.Akerlof$^{\dagger,\spadesuit}$,
R.A.  Allsman$^\ast$, \
T.S. Axelrod$^\ast$, }
\centerline {D.P. Bennett$^{\ast,\dagger}$, S. Chan$^\ddagger$,
K.H. Cook$^{\ast,\dagger}$, K.C. Freeman$^\ddagger$,}
\centerline {K. Griest$^{\dagger,\|}$, S.L. Marshall$^{\dagger,\flat}$, H-S.
Park$^\ast$, \
S. Perlmutter$^\dagger$,}
\centerline {B.A. Peterson$^\ddagger$, M.R. Pratt$^{\dagger,\flat}$, P.J.
Quinn$^\ddagger$, \
A.W. Rodgers$^\ddagger$,}
\centerline {C.W. Stubbs$^{\dagger,\flat}$, W. Sutherland$^\dagger$}
\centerline { (The MACHO Collaboration) }}
\centerline {CfPA Preprint--93--30: to appear in {\it Nature}}
\vskip 0.5truein
\centerline{$^\ast$ Lawrence Livermore National Laboratory, Livermore, CA
94550}
\vskip 8pt
\centerline{$^\dagger$ Center for Particle Astrophysics, University of
California, \
Berkeley, CA 94720}
\vskip        8pt
\centerline{$^\ddagger$ Mt.  Stromlo and Siding Spring Observatories,}
\centerline{Australian National University, Weston, ACT 2611, Australia}
\vskip 8pt
\centerline{$^\flat$ Department of Physics, University of California, \
Santa Barbara, CA 93106 }
\vskip 8pt
\centerline{$^\|$ Department of Physics, University of California, \
San Diego, CA 92039 }
\vskip 8pt
\centerline{$^\spadesuit$  Department of Physics, University of Michigan, Ann
Arbor,
MI 48109}

\vskip 0.2truein
\centerline{ \lastrev }   
\vskip 1.0truein

\parskip 0pt  

{\bf
There is now abundant evidence for the presence of large quantities
of unseen matter surrounding normal galaxies, including our own$^{1,2}$. The
nature of this `dark matter' is unknown, except that it cannot be made
of normal stars, dust, or gas, as they would be easily detected.
Exotic particles such as axions, massive neutrinos or other weakly
interacting massive particles (collectively known as WIMPs)
have been proposed)$^{3,4}$, but have yet to be detected. A less exotic
alternative is normal matter in the form of bodies with masses ranging from
that of a large planet to a few $ \msun$. Such objects, known
collectively as massive compact halo objects$^5$ (MACHOs) might be
brown dwarfs or `Jupiters' (bodies too small to produce their own
energy by fusion), neutron stars, old white dwarfs, or black holes.
Paczynski$^6$ suggested that MACHOs might act as gravitational
microlenses, occasionally causing the apparent brightness of distant
background stars temporarily to increase. We are conducting a
microlensing experiment to determine whether the dark matter halo of our
galaxy is made up of MACHOs. Here we report a candidate for  a
microlensing event, detected by monitoring the light curves of 1.8 million
stars in the Large Magellanic Cloud for one year.  The light curve shows
no variation for most of the
year of data taking, and an upward excursion lasting over 1 month,
with a maximum increase of $\approx \bf 2$ mag. The most probable lens
mass, inferred from the duration of the event, is $\bf \sim 0.1 \,\msun$.
} 

The MACHO Project$^{7,8}$ uses the gravitational microlens
signature
to search for evidence of MACHOs in the Galactic halo, which is thought
to be at least three times more massive than the visible disk$^2$.
(Two other groups are attempting a similar search$^{9,10}$.)
If most of our Galaxy's dark matter resides in MACHOs,
the `optical depth' for microlensing towards the Large Magellanic Cloud (LMC)
is about $5 \ten{-7}$ (independent of the mass function of MACHOs),
 so that at
any given time about one star in two million
will be microlensed with an amplification factor $\rm A > 1.34$
(ref.~5).
Our survey takes advantage of the transverse motion of MACHOs
relative to the line-of-sight from the observer to a background star.
This motion causes a transient, time-symmetric and
achromatic brightening that is quite unlike
any known variable star phenomena, with a characteristic timescale
$\hat{t} = 2 r_E / v_{\perp}$  where
$r_E$ is the Einstein ring radius and $v_{\perp}$ is the MACHO
velocity transverse to the line-of-sight.
 For typical halo models, the time with  $\rm A > 1.34
 \sim 100 \sqrt{\rm M_{macho}/\msun}$ days$^5$ (where
$\msun$ is the mass of the sun).
The amplification can be large, but these events
are extremely rare; for this reason our survey was designed to follow
$ > $ ten million stars over several years.

The survey employs a dedicated 1.27m telescope  at Mount~Stromlo.
A field-of-view of 0.5 square degrees is achieved
by operating at the prime focus.
The optics include a dichroic
beamsplitter which allows simultaneous imaging in a `red' beam
($6300 - 7600\angs$)
and a `blue' beam
($4500 - 6300 \angs$).
Two large charge coupled device (CCD) cameras$^{11}$
are employed at the two foci;
each contain a $2 \times 2$ mosaic of $2048 \times 2048$  pixel
Loral CCD imagers.
The 15 $\mu$m pixel size corresponds to $0.63$ arcsec
on the sky.  The images are read out through a 16 channel system, and
written  into dual ported memory in the data acquisition computer.
Our primary target stars are in the LMC. We also
monitor stars in the Galactic bulge and the Small Magellanic Cloud.
As of 15 September 1993, over $12000$
images have been taken with the system.

The data are reduced with a crowded-field photometry routine
known as Sodophot, derived from Dophot$^{12}$.
First, one image of each field that was obtained in good seeing
is reduced in a manner similar to
Dophot to produce a `template' catalog of star positions and magnitudes.
Normally, bright stars are matched with the
template and used to determine an analytic point spread function (PSF) and
a coordinate transformation.  Photometric fitting is then performed on
each template star in descending order of brightness,
with the PSF for all other stars subtracted from the frame.
When a star is found to vary significantly, it and its neighbors undergo
a second iteration of fitting.
The output consists of magnitudes and  errors for the two colors,
and six additional useful parameters (such as the $\chi^2$ of the PSF fit and
crowding information).  These are
used to flag questionable measurements that arise from cosmic ray events
in the CCDs, bad pixels, and so on.

These photometric data
are subjected  to an automatic time-series analysis which
uses a set of optimal filters to search for microlensing candidates
and variable stars (which we have detected in abundance$^{13}$).
For each microlensing candidate a  light-curve is fitted,
and the final selection is done automatically using criteria (for example
signal-to-noise, goodness of fit, achromaticity, color)
that were established empirically using Monte Carlo
addition of fake events into real light-curves.

This analysis has been done on four fields near the center of the LMC,
containing 1.8 million stars, with approximately 250
observations for each star.
The candidate event reported here occurs in the light-curve of a star at
coordinates $\alpha = 05\ 14\ 44.5,  \delta = -68\ 48\ 00$
(J. 2000). (A finding chart is available on request from C.~A.).
The star has median magnitudes $V \sim 19.6$, $R \sim 19.0$, consistent
with a clump giant (metal-rich helium core burning star) in the LMC.
These magnitudes are
estimated using color transformations from our filters to $V$ and $R$ that
have been derived from observations of standard stars.

Our photometry for this star, from July 1992 to July 1993
is shown in Figure~1, and the candidate event
is shown on an expanded scale in Figure~2, along with the
color light curve.  The color changes by
$< 0.1$  mag as the star brightens and fades.
A mosaic, showing portions of some of the CCD images
used, is shown in Figure~3, with the relevant star at the center.
The integrated number of PSF photoelectrons detected
above the sky background in the template image is
$\approx 10^4$, for a 300 sec exposure.
The increase in counts during the
peak is highly significant, as is clear from
the Figures.

Also shown in Figure 2 is a fit to the theoretical microlensing
light-curve (see ref. 6). The four parameters fit are (1) the baseline
flux; (2) the maximum amplification $A_{max} = 6.86 \pm 0.11$;
(3) the duration $\hat{t} = 33.9 \pm 0.26$ days;
(4) the centroid in time $433.55 \pm 0.04$ days. The quoted errors
are formal fit errors.
Using the PSF fit uncertainties as
determined by the photometry program, the best-fit microlensing curve
gives a $\chi^2$ per degree of freedom of 1.6 (for 443 d.o.f.).

A number of features of the candidate event are consistent with
gravitational microlensing: the light curve is
achromatic within measurement error, and it has the expected
symmetrical shape.
If this is a
genuine microlensing event, the mass of the deflector can be estimated.
Since the duration depends upon
the lens mass, the relative velocity transverse to the line-of-sight,
and the distance to the lens, (none of which are known),
the lens mass cannot be uniquely determined from the
duration.
However, using a model of the mass and velocity distributions of halo
dark matter,
one can find the relative probability that a MACHO of
mass $m$ gave rise to the event.
Thus, if this is geniune microlensing, Fig. 9 of Ref. 5 implies the
most likely mass is approximately $0.12 \msun$,
with masses of $0.03 \msun$ and $0.5 \msun$ being roughly half as likely.
However, this method does not properly take into account our
detection efficiencies, and should be considered only a rough estimate.

The mass range given above includes brown dwarfs and main sequence stars.
Any microlensing star is very unlikely to be a red dwarf of
the Galactic stellar halo, because one can show that the optical depth $\tau_*$
for microlensing by main sequence stars of the stellar halo is very low.
Even if the mass function of the stellar halo rises as steeply as
$dN/dM \propto M^{-4}$, as suggested recently$^{14}$ (here N is
the number of stars per unit stellar mass interval), $\tau_*$ is still a few
hundred times smaller than the $5 \times  10^{-7}$ optical depth
estimated for MACHO
microlensing. The chance of finding such a stellar microlensing event
among our 1.8 million stars is therefore very small.

The prospects for direct observation of a lensing object are not
favorable.  Even a star of $0.5 \msun$, for example,
would have $V \sim 24$, and for many years would be within
a small fraction of an arcsecond of the much brighter LMC star.

We emphasize that the observed  stellar brightening could
be due to some previously unknown source of intrinsic
stellar variability.
The fit discrepancy near the peak is not yet understood; a more
refined analysis of the data is underway.
We do not yet have a spectrum
of the star.
A crucial test of the hypothesis that we are seeing gravitational
microlensing by MACHOs in the galactic halo will be the detection
of other candidates.
So far, we have analysed only $\sim 15\%$ of our first
year's frames, and we plan to continue observations until 1996;
this will allow us to determine whether or not gravitational microlensing
is really the cause.
Additional events should show the theoretically expected distribution of
maxima,
and should be representative of both the color-magnitude diagram and the
spatial structure of
the LMC.  No repeats should be seen in any given star!
(While this paper was in preparation, we were informed by
J.~Rich (personal communication) of the candidate events
reported by the EROS collaboration.
Note that the two
groups use different definitions of characteristic time.)

If such candidates do result from microlensing we should be
able to determine the
contribution of MACHOs to the dark matter in the Galactic halo.
The results presented here encourage us to believe this will happen.

\vskip 14pt

We are very grateful for the skilled support given our project
by the technical staff at the Mt. Stromlo
Observatory.
Work performed at LLNL is supported by the DOE under contract W7405-ENG-48.
Work performed by the Center for Particle Astrophysics on the UC campuses
is supported in part by the Office of Science and Technology Centers of
NSF under cooperative agreement AST-8809616.
Work performed at MSSSO is supported by the Bilateral Science
and Technology Program of the Australian Department of Industry, Technology
and Commerce. KG acknowledges a DOE OJI grant, and CWS thanks the Sloan
Foundation for their support.

\vfill\eject

\sect{Figure Captions}

{\parskip 10pt  %
\fig{Figure 1}
The observed light-curve
with estimated $\pm 1\sigma$ errors.
The upper panel shows $A_{blue}$, the flux (in linear units)
divided by the median
observed flux, in the blue passband.
The lower panel is the same for the red passband.
The smooth curve shows the
best-fit theoretical microlensing model, fitted
simultaneously to both colors.
Time is in days from 1992 Jan~2 00:00 UT.

\fig{Figure 2} As in Figure~1, with expanded scale around the event.
The curve shows the
theoretical microlensing model, fitted
simultaneously to both colors.
The bottom panel is the color light curve, showing the ratio of
red to blue flux, normalized so that the median is unity.

\fig{Figure 3} Selected Red CCD frames
centered on the microlens candidate, showing
observations before,
during and after the event.

 }  

\sect{References}
\parskip 0pt

\ref 1. Trimble, V.
{\it Ann.  Rev.  Astron.  Astrophys.}, {\bf 25}, 425--472 (1987).

\ref 2. Fich, M.  \& Tremaine, S.
{\it Ann.  Rev.  Astron.  Astrophys.}, {\bf 29}, 409--445 (1991).

\ref 3. Primack, J.R.,  Seckel, D. \& Sadoulet, B.
  {\it Ann.  Rev.  Nucl. Part.  Sci.}, {\bf B38}, 751--807 (1988).

\ref 4. Kolb, E.W.  \& Turner, M.S. {\it `The Early
Universe'}, Addison Wesley:  New York (1990).

\ref 5. Griest, K. {\it Astrophys. J.}, {\bf 366}, 412--421 (1991).


\ref 6. Paczynski, B. {\it Astrophys. J.}, {\bf 304}, 1--5 (1986).

\ref 7.  Alcock, C. \etal {\it Astron. Soc. Pacific Conf. Series} {\bf 34},
   193--202 (1992).

\ref 8. Bennett, D. \etal {\it Ann. N.Y. Acad. Sci.},
    {\bf 688}, 612--618 (1993).

\ref 9. Magneville, C. {\it Ann. N.Y. Acad. Sci.},
    {\bf 688}, 619--625 (1993).


\ref 10. Udalski, A. \etal {\it Ann. N.Y. Acad. Sci.}, {\bf 688},
      626--631 (1993).

\ref 11. Stubbs, C. \etal {\it SPIE Proceedings} {\bf 1900}, 192--204 (1993).

\ref 12. Schechter, P.L., Saha, A. \& Mateo, M.L.
  {\it Pubs. Astron. Soc. Pacific}, in press.

\ref 13. Cook, K.H.  \etal, {\it BAAS}, {\bf 24}, 1179 (1993).

\ref 14.  Richer, H.B. and Fahlman G.G.,  Nature, 358 383 (1992).

\vfill
\eject
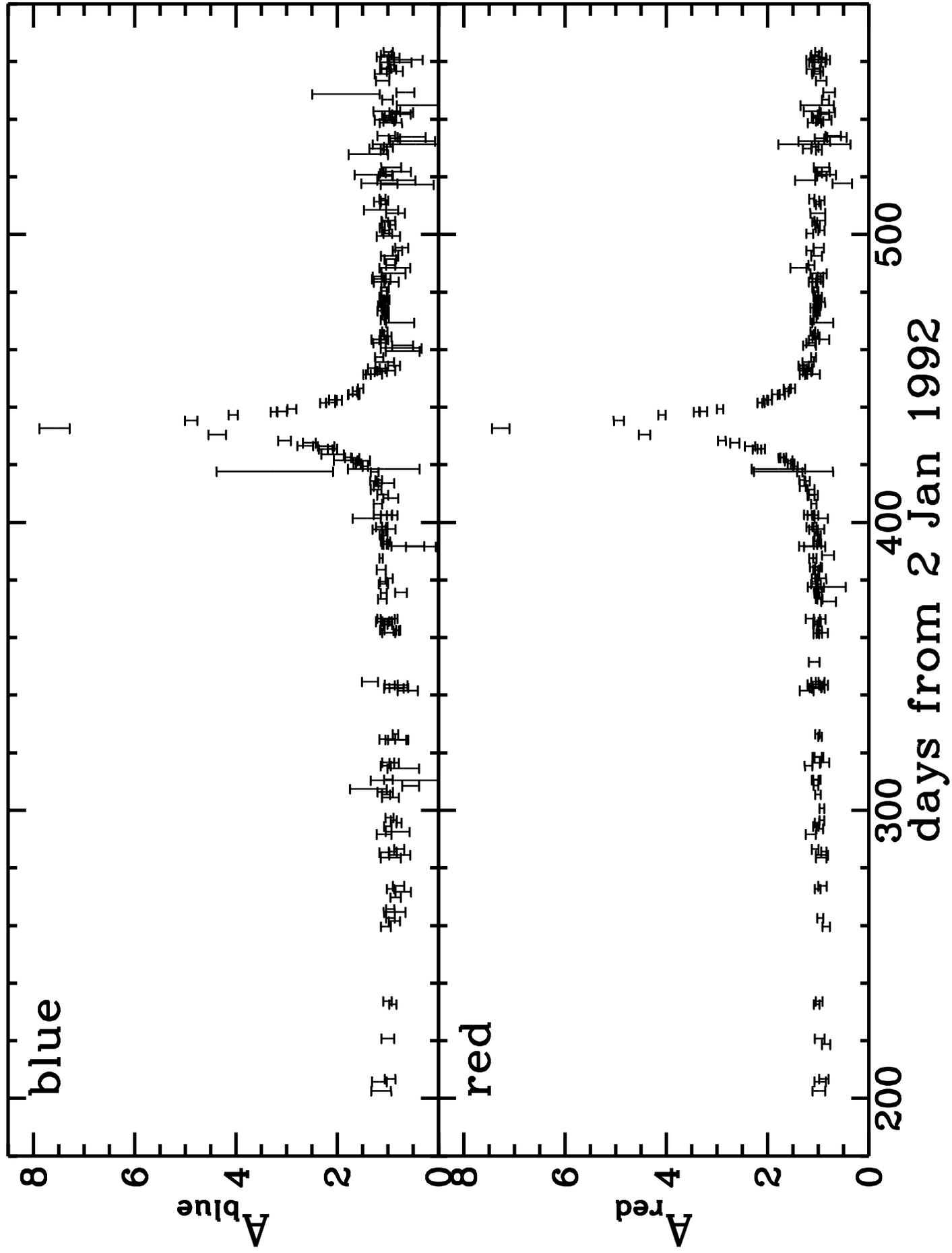
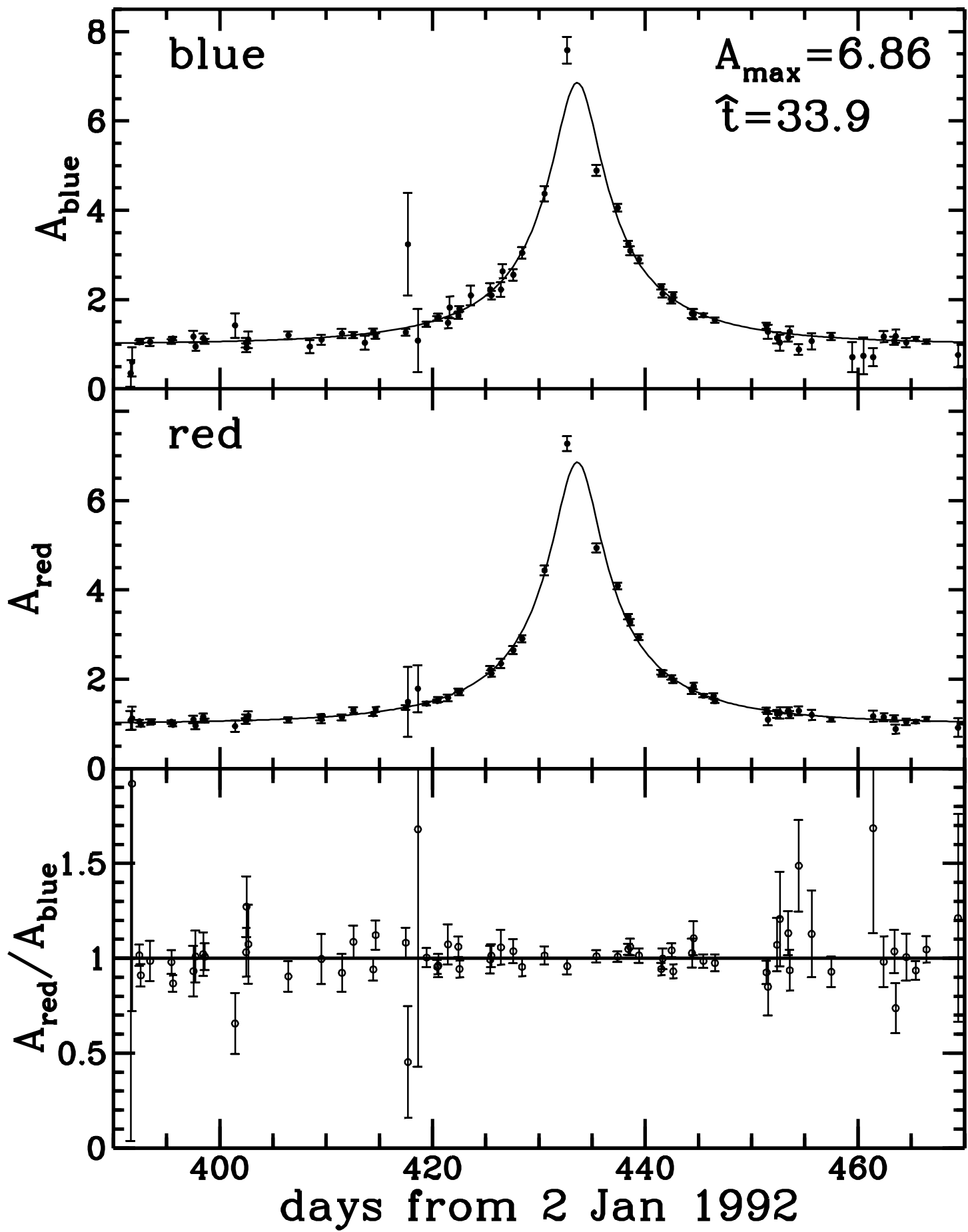
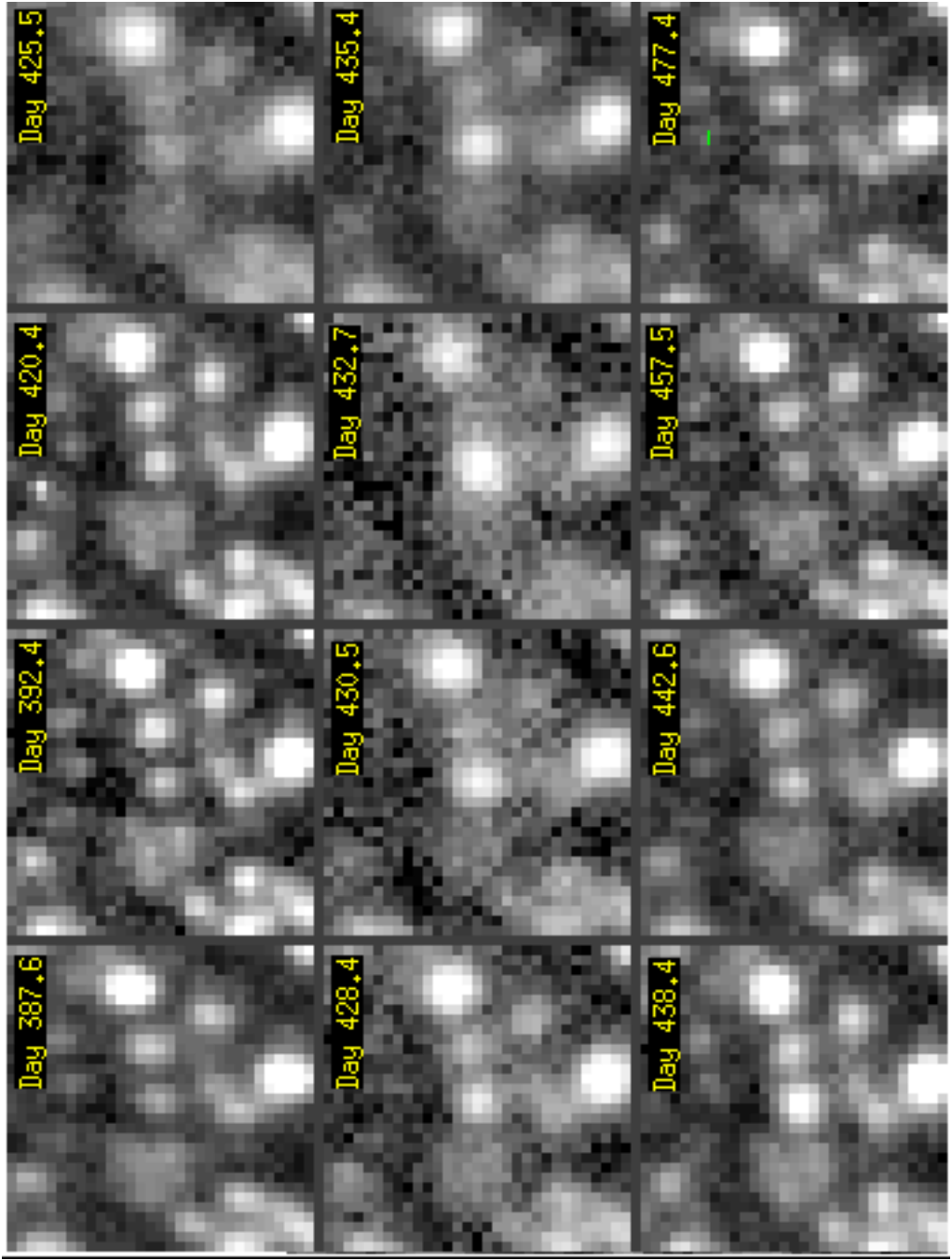
\end